\documentclass[a4paper,11pt]{article}
\usepackage{graphicx} 

\pdfoutput=1 

\usepackage{jheppub} 

\usepackage{array}

\usepackage{mathrsfs}
\usepackage{comment}
\usepackage{centernot}
\usepackage{mathtools}
\usepackage{stmaryrd}

\usepackage{yfonts}

\usepackage{latexsym,amssymb,amstext,amsmath}
\usepackage{hyperref}
\usepackage{graphbox,graphicx}
\usepackage{amsthm}
\usepackage{esint}
\usepackage{floatrow}

\usepackage{footmisc}

\def\bea {\begin{eqnarray}}
\def\eea {\end{eqnarray}}

\def\beq{\begin{equation}}
\def\eeq{\end{equation}}
\def\beqa{\begin{eqnarray}}
\def\eeqa{\end{eqnarray}}

\theoremstyle{definition}

\title{\boldmath 
Holography from number theory: Emergent holographic AdS$_2$ space-time from exact BPS black hole microstate counting}

\author{Gabriel Lopes Cardoso${}^1$ and Suresh Nampuri${}^2$}

\affiliation{${}^1$ Center for Mathematical Analysis, Geometry and Dynamical Systems,\\
Department of Mathematics, Instituto Superior T\'ecnico, Universidade de Lisboa,\\
Av. Rovisco Pais, 
1049-001 Lisboa, Portugal}
\affiliation{${}^2$ Center for Mathematical Studies,	
Faculdade de Ci\^encias, Universidade de Lisboa,\\
Edf. C6, 
Campo Grande, 1749-016 Lisboa, Portugal }

\emailAdd{gabriel.lopes.cardoso@tecnico.ulisboa.pt}
\emailAdd{nampuri@gmail.com}

\abstract{In this note, starting from the exact counting formulae for 4D BPS black hole degeneracies in the cases of 1/2 BPS $\mathcal{N}=4$ and 1/8 BPS $\mathcal{N}=8$ solutions, expressed as Rademacher expansions for coefficients of modular and Jacobi forms respectively, we explicitly show how the number theoretic data in each Rademacher summand is encoded by a Segal-Bargmann heat kernel calculation in the de Alfaro-Fubini-Furlan (DFF) model of conformal quantum mechanics, thus demonstrating the holographic origin of the BPS Bekenstein-Hawking entropy, and all logarithmic and power law suppressed corrections to it. We then derive the 2D holographically dual bulk space-time for this CQM and show that it is an AdS$_2$ space-time with radial length scale correlated to the energy scale of the DFF model. We further show how the Bekenstein-Hawking entropy emerges from an entanglement entropy computation in this bulk space-time. Hence, for the two classes of 4D BPS black holes under consideration, we have inferred the holographic CQM relevant for microscopic state counting from the exact number-theoretic degeneracy formulae and demonstrated the emergence of the near-horizon AdS$_2$ attractor geometry that is holographically dual to it.
 
 }

\begin{document}

\maketitle

\flushbottom

\section{Introduction}

BPS black holes in asymptotically flat space-time occupy a unique place in string theory as a mathematically consistent theory of quantum gravity. For one, as has recently been suggested by \cite{McSharry:2025iuz}, BPS black holes cannot be seen as the natural endpoint of the radiation process of non-extremal black holes, and hence they represent zero temperature states in gravity that cannot be approached in a finite number of steps from a non-zero temperature state.
 This implies the statistical description of their microstate degeneracies will be disjunct from that of non-supersymmetric black holes. Fortuitously, exact expressions for microstate degeneracies have been computed precisely for some classes of BPS black holes in asymptotically flat space-time, by modeling them in terms of bound states of strings and branes \cite{Sen:1995in,Dabholkar:1995nc,Maldacena:1999bp,Dabholkar:2005dt}.  These expressions are constructed from automorphic forms of negative weight, and the degeneracies in a given model are obtained via a Rademacher series expansion. Typically, each term in this expansion involves a modified Bessel function of the first kind ${\hat I}_{a} (z)$,
where $z$ is expressed in terms of the black hole data. The modified Bessel function contains both the leading contribution to the exact BPS degeneracy as well as the subleading power law suppressed and logarithmic corrections to it. 

In \cite{LopesCardoso:2025azr}, it was shown that this Bessel function stripped of its exponential factor, coincides with the Euclidean heat kernel $K_E(x_0, x_0, T)$
of the DFF$_\omega$ model for periodic paths that begin and end at $x_0$, where $x_0$ is expressed in terms of the 
two couplings of the DFF$_\omega$ model, $g$ and $\omega$.
The DFF$_\omega$ model is the conformal quantum mechanics model (CQM) of de Alfaro, Fubini, and Furlan with an additional harmonic trapping potential \cite{deAlfaro:1976vlx,Chamon:2011xk,Jackiw:2012ur}, and it possesses an  $SL(2, \mathbb{R})$ symmetry\footnote{In an AdS$_2$ space-time, its three generators, $H, R, S$, generate Euclidean time translations in Poincar\'e patch time, black hole time and global time, respectively, see \cite{LopesCardoso:2025kry}.}.
This identification of ${\hat I}_{a} (z)$
with $K_E(x_0, x_0, T)$ arises by expressing both $a$ and $z$ in terms of $g$ and $\omega$, and subsequently taking the small $\omega$ limit. From the perspective of the AdS$_2$/CQM correspondence, this suggests that the DFF$_\omega$ model describes a universal sector of the holographic conformal quantum mechanics dual to the near-horizon AdS$_2$ geometry of the BPS black hole, which can be formulated in terms of a DFF$_{\omega}$ model restricted to a specific state $|0\rangle$. In this note, we will build upon these results to develop a bottom-up (boundary to bulk) approach to holography via the known number theoretic formulae for exact BPS black hole degeneracies.

Our starting point 
 is motivated by \cite{Gibbons:1998fa}, which proposed a microscopic description of the extremal Reissner–Nordstr\"om black hole in the near-horizon limit, based on the large $N$ limit of the $N$-particle rational Calogero (super) conformal quantum mechanics model. It was shown in \cite{Lechtenfeld:2015wka} that in the $N \rightarrow \infty$ limit, the state degeneracy at large energy $M$ grows exponentially as $e^{2 \pi \sqrt{M/6}}$,
and this was used in \cite{LopesCardoso:2025azr} to argue that the leading contribution $e^z$ 
to the exact BPS black hole degeneracy should arise from the large $N$ limit of the $N$-particle rational Calogero model. This suggests the Calogero model as a viable starting point for deriving a holographic quantum mechanical description. 

Therefore, we first revisit the $N$-particle rational Calogero (super) conformal quantum mechanics model and show 
that the Bessel function ${\hat I}_{a} (z)$ can be traced back to the large $N$ Calogero system. Namely, we 
demonstrate how, in the $N \rightarrow \infty$ limit, the $N$-particle rational Calogero model can be related to a DFF$_\omega$ model with couplings $g_{\rm eff}, \omega$, evaluated on a particular quantum state. This 
 state is the vacuum $|0\rangle$ of the $SL(2, \mathbb{R})$ generator $R$, whose wave function becomes sharply  peaked, in the semiclassical limit $\hbar \rightarrow 0$, at 
 $x_0 = \left( 2 g_{\rm eff}/m \omega^2 \right)^{1/4}$.
In this way, one sees explicitly how the Euclidean heat kernel $K_E(x_0, x_0, T)$ 
emerges from the large $N$ dynamics of the rational Calogero model.

The $N = \infty$ limit of the  $N$-particle rational Calogero model can be understood thus. 
As $N \rightarrow \infty$, the dynamics reorganizes into a collective field theory: the discrete set of particles \{$x_i$\} is replaced by a continuous density $\rho(x)$, with $\rho(x)$ and its conjugate momentum forming the fundamental degrees of freedom.  The resulting field theory contains a distinguished collective mode that obeys an effective conformal quantum mechanics of the DFF$_\omega$ type.
Hence,
the $N=\infty$ limit is best understood as a collective field theory, whose  radial sector reduces to a  DFF$_\omega$ model in the $SL(2,\mathbb{R})$  vacuum state $|0\rangle$\footnote{The DFF model and 
the $N$-particle rational Calogero model are also known to have a matrix model origin \cite{Polychronakos:1997nz,Verlinde:2004gt,
Polychronakos:2006nz,Masuku:2015vta}, which points to a matrix model description of BPS black hole entropy.
}. We compute the associated Euclidean heat kernel $K_E (x_0, x_0, T)$ in this model. 
We then demonstrate how the modified Bessel function ${\hat I}_a (z)$ that encodes the Bekenstein-Hawking entropy as well as the logarithmic and power law suppressed correction terms in each Rademacher summand, is captured by an amputated heat kernel of the DFF$_{\omega}$ model, calculated in a complexified configuration space.
This constitutes the first known holographic derivation of BPS black hole entropy from a candidate holographic CQM dual to the near-horizon AdS$_2$ geometry. We subsequently explicitly demonstrate the holographic encoding of space-time by this CQM by deriving an emergent holographically dual Poincar\'e AdS$_2$ space-time, whose radial coordinate encodes the energy scale of the CQM. This is done as follows.

The vacuum state $|0\rangle$ is not an eigenstate of the Hamiltonian $H$, but of the generator $R$
and diffuses under time evolution in $H$ across the $R$-spectrum. 
We analyse the temporal diffusion of the vacuum state $|0\rangle$ in the Krylov basis, wherein said diffusion is formulated in terms of the time evolution of the Lanczos diffusion coefficients, expressed as a hopping equation between sites of a discrete lattice, where the sites are indexed by the integer that counts discrete energy levels in the spectrum. The continuum limit of this discrete equation yields a coordinate $r$ that represents an energy scale of the quantum mechanical system. The discrete equation then yields a time evolution Schr\"odinger equation in the continuum variables $t$ and $r$.

This Schr\"odinger equation is invariant under an $SL(2,\mathbb{R})$ symmetry group. We geometrise this symmetry group by demanding it to be the isometry group near the boundary $r=0$ of the AdS$_2$ metric in Poincar\'e coordinates $(t,r)$,
whose radial coordinate corresponds to the energy scale of the DFF$_\omega$ conformal quantum mechanics (CQM) model. Hence, this metric is a candidate metric for a holographic bulk dual to the CQM under the AdS$_2$/CQM correspondence. Therefore, we identify this metric with the AdS$_2$ near-horizon attractor geometry, which is an exact solution of a 2D theory of gravity, obtained by a spherical reduction of the relevant parent 4D low-energy effective superstring action, whose equations of motion support the BPS black hole background as a solution. The 2D gravity theory contains a negative cosmological constant and is by itself known to be a chiral 2D CFT \cite{Strominger:1998yg}. 
We determine its central charge by reading it off from the Schwarzian term arising from a conformal transformation of the Euclidean Poincar\'e patch coordinates $(t_E, r)$ to the black hole coordinates ($\rho,\varphi$) and equate it to the vacuum expectation value of the $R$ operator, which generates translations in the $\varphi$ coordinate. 
This enables us to define the vacuum entanglement entropy\footnote{Our approach differs from that of \cite{Azeyanagi:2007bj} who compute entanglement entropy in global AdS$_2$ via the method of conical defects, as well as \cite{LopesCardoso:2025kry} which computes the entanglement entropy in a thermofield double formalism in global AdS$_2$.}
 of the chiral 2D CFT 
of a finite segment of length $L$ on 
a $x= \ln  r$ coordinate as  ${\cal S}_{\text{ent}}= \frac{c}{6} \ln (\frac{L}{\epsilon})$ \cite{Holzhey:1994we} , which by a suitable transformation to the thermal cylinder, reproduces the leading order BPS Bekenstein-Hawking entropy as the thermal entropy in the high temperature limit.

In sum, starting from the number-theoretic formula for the BPS degeneracy, we demonstrate the encoding of each of its summands (upto phase factors) in terms of a diagonal  Segal-Bargmann heat kernel  
of the DFF$_\omega$ CQM.  We then explicitly demonstrate how the AdS$_2$  attractor geometry that is the holographic bulk dual to the CQM, emerges from the CQM. The AdS$_2$/CQM duality between the radial scale in the bulk and the energy scale of the holographic CQM is intrinsic to this construction. Subsequently we compute entanglement entropy in the bulk for a segment along the radial coordinate $x = \ln \,r$ via the standard vacuum entanglement entropy formula and show how this encodes the Bekenstein-Hawking entropy. This showcases the Bekenstein-Hawking entropy as an entanglement entropy in the Poincar\'e AdS$_2$. 
Note that this bulk entanglement entropy is in fact an entanglement entropy on a energy scale of the holographic CQM and is therefore equally appropriately labeled as a renormalisation entropy \cite{Holzhey:1994we}.

We apply the above formalism to the case of 
$1/2$ and $1/8$ BPS black holes in $4D$ toroidally compactified ${\cal N}=4$ heterotic string and ${\cal N}=8$ type IIB superstring theory, whose 
exact BPS microstate degeneracies are encoded in modular and Jacobi forms, respectively.

The paper is organized as follows. 
In Section \ref{sec:dff} we show how the $N$-particle rational Calogero conformal quantum mechanics model gives rise to a DFF$_\omega$ model,
evaluated on a particular quantum state (denoted $|0\rangle$) in the $N \rightarrow \infty$ limit. 
In Section \ref{sec:scl} we review salient features of the DFF$_\omega$ model and we discuss its semi-classical limit. In Section \ref{sec:heat} we discuss the Segal-Bargmann heat kernel in the DFF$_\omega$ model and exhibit its relation with the Rademacher series that give the exact BPS microstate degeneracies. In Section \ref{sec:kloo} we discuss how to reproduce the Kloosterman sums that appear in the Rademacher series from a quantum mechanics point of view. In Section \ref{sec:Kryl} we analyze the temporal evolution of the state $|0\rangle$ in the Krylov basis. Taking a continuum limit, we obtain a Schr\"odinger equation with an $SL(2, \mathbb{R})$ symmetry. By geometrising the latter, we obtain an $AdS_2$ metric in Poincar\'e coordinates, whose radial coordinate corresponds to the energy scale in the DFF$_\omega$ model. By viewing the associated 2D gravity theory as a chiral CFT$_2$, we deduce its central charge. We end with a summary of our results in Section \ref{sec:conc}. In Appendix \ref{sec:appA} we relate the leading BPS black hole entropy to the vacuum entanglement entropy of a finite segment on an infinite line. In Appendix \ref{appB} we relate the emergent  $AdS_2$ metric in Poincar\'e coordinates to a Fisher information metric. In Appendix \ref{sec:AppC} we discuss the Hamiltonian origin of the multiplier system in the Rademacher series of $1/8$ BPS black holes in ${\cal N}=8$ type IIB supersymmetric string theory.

\section{From the microscopic Calogero model to the holographic DFF$_\omega$ model \label{sec:dff}}

In \cite{Gibbons:1998fa}, it was argued that the microscopic modeling of BPS black hole degeneracy in terms of $N\gg 1$ BPS probe particles in the AdS$_2$ near-horizon geometry yields a putative holographic quantum mechanics\footnote{For more on how conformal actions emerge in the context of near-horizon AdS$_2$,
cf. \cite{Michelson:1999dx,Cacciatori:1999rp,Cadoni:2000gm}.}. The Hamiltonian in this case is given by that of the $N$-particle Calogero model in a harmonic trap, 
\beq\label{Cal}
H_{\text{Cal}} = \sum_{1\leq i \leq N}\left( \frac{p_i^2}{2m} + \left(\sum_{i<j\leq N}\frac{g}{|x_i-x_j|^2}\right)+ \frac{1}{2} m \omega^2 x_i^2 \right) \;\;\;,\;\;\; g> 0 \;.
\eeq
In the large $N$ limit, the state degeneracy $p(M)$ at energy level $M$ exhibits an asymptotic exponential growth \cite{Lechtenfeld:2015wka}. Assuming $s$ different, mutually non-interacting species of Calogero particles, this asymptotic growth is $e^z$, with $z= 2 \pi \sqrt{\frac{M s}{6}}$, and captures the
 Bekenstein-Hawking entropy of BPS black holes for suitable values of $s$ \cite{LopesCardoso:2025azr}.

We now show how in the large $N$ limit, this model reduces to a single DFF degree of freedom coupled to a harmonic trap evaluated on a static state, i.e. with zero momentum. This static state corresponds to an extremum of the potential in this DFF model. 

For the purpose of computing microscopic degeneracies, we extract out the centre of mass (CM) degree of freedom to obtain the relevant Hamiltonian capturing the relative degrees as 
\bea
H_{\text{rel}} &=& \sum_{1\leq i  \leq N}\left( 
\frac{\pi_i^2}{2m} + \left(\sum_{i< j\leq N} \frac{g}{|x_i-x_j|^2}\right)+ \frac{1}{2} m \omega^2 x_i^2 \right) \;,\,\nonumber\\
\sum_{1\leq i\leq N}{x_i}&=&\sum_{1\leq i\leq N}{\pi_i}=0.
\eea
In the above, $x_i$ represent co-ordinates in \eqref{Cal}  shifted relative to the CM, though we retain the same symbols for notational simplicity, while the $\pi_i$ represent the momenta conjugate to the $N-1$ relative degrees of freedom, which can, in turn be formulated in spherical co-ordinates as a collective radial and $N-2$ angular degrees of freedom. For an eigenfunction $\psi(r,\Omega)$ of the associated quantum Hamiltonian ${\hat H}_{\text{rel}}$ satisfying 
\bea {\hat H}_{\text{rel}} \psi &=& E \psi, 
\eea
the variable separability $\psi(r,\Omega)= R(r)Y(\Omega)$ gives rise to the  
radial Hamiltonian operator,
\bea
{\hat H}_{\text{radial}}&=& - \frac{\hbar^2}{2m} \left( 
\partial_r^2+ \frac{(N-2)}{ r}  \partial_r \right) + \frac{\lambda_k}{r^2} + \frac{1}{2} m \omega^2 r^2,
\eea  
where
\bea \lambda_k = \frac{\hbar^2}{2m} \,  l_k (l_k+N-3)
\eea 
and 
 \bea l_k = k + \frac14 \left(1 +  \sqrt{1+\frac{8 m g}{\hbar^2}}\right) N(N-1) \;\;\;,\;\;\; k\in \mathbb{N}_0 \;,\eea
with $l_k$ denoting the angular momentum quantum number corresponding to the $k$-th eigenvalue of the angular operator.

Setting $R(r) =  r^{-\frac{N-2}{2}} {\tilde R}(r)$,
we eliminate the single derivative term in the radial Hamiltonian operator, to obtain an effective DFF Hamiltonian operator coupled to a harmonic oscillator as 
\bea 
{\hat H}_{\text{DFF}_\omega}= \frac{p_r^2}{2m} + \frac{1}{r^2} \left(\lambda_k - \frac14 (N-3)^2 + \frac14 \right)  + \frac{1}{2} m \omega^2 r^2, 
\eea

Now, at large $N$, $\lambda_k$ behaves as $g_{\rm eff} N^4$, where   
\bea
g_{\rm eff} =\frac{\hbar^2}{2m}\frac{(1+\sqrt{1+\frac{8 m g}{\hbar^2}})^2}{16} \;.
\eea
Therefore, using the large $N$ scaling where $r \rightarrow N^{\frac{3}{2}} r$ and $\omega \rightarrow \frac{\omega}{N}$, we obtain 
\bea 
{\hat H}_{\text{DFF}_\omega} = \frac{p_r^2}{2m N^3} + N \left( \frac{g_{\rm eff}}{ r^2}   + \frac{1}{2} m \omega^2 r^2 \right).
\eea We can easily see therefore that the large $N$ limit picks out the zero radial momentum state.\footnote{The CM Hamiltonian, $H_{\text{CM}}= \frac{P_{\text{CM}}^2}{2mN}+ \frac{1}{2} N m \omega^2 X_{\text{CM}}^2$, is also led to operate on a static state in this limit.  } The corresponding potential has an overall $N$ scaling which can be absorbed into the Euclidean time periodicity $\beta$ when evaluating the action. We then require this rescaled periodicity to be finite in the large $N$ limit.

Hence, to sum up, in the large $N$ limit, the radial mode dynamics of the Calogero model, with fixed angular momentum $l_k$, is captured by $N$ independent identical DFF degrees of freedom, each coupled to a harmonic oscillator, in the vacuum state with zero momentum. Hence, starting from a model of BPS black hole degeneracies as encoded by BPS particles in the near-horizon geometry \cite{Gibbons:1998fa}, we have demonstrated how the DFF model evaluated on a quantum state, whose semiclassical limit is the zero momentum state, could emerge in the microstate counting formula as a putative holographic picture for BPS black hole entropy.  

In Section \ref{sec:heat} we will 
demonstrate how this model encodes number-theoretic data in the BPS degeneracy formula, expressed as a Rademacher expansion, in the 1/2 BPS $\mathcal{N}=4$ and 1/8 BPS $\mathcal{N}=8$ black hole cases.

\section{The semi-classical limit in the DFF$_\omega$ model \label{sec:scl}}

The DFF model, a conformal quantum mechanics model \cite{deAlfaro:1976vlx,Chamon:2011xk,Jackiw:2012ur} with a coupling constant $g>0$, and a Hamiltonian (restricted to the half-line $x >0$) given by
\bea
H =  \frac{p^2}{2m} + \frac{g}{x^2}  \;\;\;,\;\;\; g > 0 \;\;\;,\;\;\; x > 0 \;,
\label{Rcl}
\eea
possesses an underlying $so(2,1)$ algebra,
whose generators $L_{\pm}, R$ satisfy
\bea
[R, L_{\pm}] = \pm L_{\pm} \;\;\;,\;\;\; [L_-, L_+] = 2 R \;.
\label{gensl}
\eea
The generator $R$ has a discrete spectrum with a unique groundstate $|0 \rangle$,
\bea
R |n\rangle =  \hbar \left( r_0 + n \right) |n \rangle \;\;\;,\;\;\; \langle n | m \rangle = \delta_{n,m} \;\;\;,\;\;\; n \in \mathbb{N}_0 \;,
\eea
where $r_0$ is expressed in terms of the coupling $g$ as 
\bea
r_0 = \frac12 \left( 1 + a \right)  \;\;\;,\;\;\; 
a = \frac12 \sqrt{1 + \frac{8 m g}{\hbar^2} } \;.
\label{rocoup}
\eea
The generator $R$ can be identified with the Hamiltonian $H'$ of the DFF$_\omega$ model, which is obtained form the DFF model
by the addition of a harmonic oscillator term $\frac12 m \omega^2 x^2$, in which case $H' = 2 \omega R$.

By sending $\hbar \rightarrow 0$, keeping $g$ fixed,
one obtains
\bea
\langle 0 | R | 0 \rangle = \hbar r_0 = \frac12 \sqrt{2 m g} + {\cal O} \left(\hbar \right)\;.
\label{semicldff}
\eea
Note that we may view \eqref{semicldff} as the expectation value of $R$ in the coherent state $|\alpha\rangle$, in the limit $\alpha \rightarrow 0$, cf. Appendix \ref{appB}. 
This defines the semi-classical limit in the DFF$_\omega$ model.
The leading term in \eqref{semicldff} is reproduced by the following classical analysis.
The equations of motion of the  DFF$_\omega$ model admit the constant solution $x_0 >0$ given by \cite{LopesCardoso:2025azr},
\bea
x_0^4 = \frac{2g}{m \omega^2} \;.
\label{x0g}
\eea
Evaluating the potential $V(x)$ of the  DFF$_\omega$ model,
\bea
V(x) = \frac{g}{x^2}  + \frac12 m \omega^2 x^2 \;,
\eea
 on the classical solution $x_0$, gives 
\bea
V(x_0) = \sqrt{2 m g } \, \omega \;.
\eea
Thus, classically, we find
\bea
R = \frac{V(x_0)}{2 \omega}
= \frac12 \sqrt{2 m g } \;,
\eea
which reproduces the semi-classical 
value \eqref{semicldff}, to leading order.

The value $x_0$ is also the approximate value around which the groundstate probability density $|\psi_0(x)|^2$ 
is peaked. The groundstate wave function reads 
\bea
\psi_0 (x) = \left( \frac{m \omega}{\hbar} \right)^{r_0} \left( \frac{2}{\Gamma (2 r_0) }\right)^{1/2} \, x^{2 r_0 - \frac12} \, e^{- \frac{m \omega}{2 \hbar} x^2} \;.
\label{normgroundw}
\eea
Extremizing the probability density
$|\psi_0(x)|^2$, one finds that $|\psi_0(x)|^2$ is peaked at $x_p$ given by
\bea
x_p^2 = \frac{(4 r_0 -1) \hbar}{2 m \omega} \;.
\eea
In the semi-classical limit $\hbar \rightarrow 0$ we have $r_0 \rightarrow \infty$, and hence $x_p$ becomes
\bea
x_p^2 \approx \frac{2 r_0  \hbar}{ m \omega} \approx \frac{\sqrt{2g} }{ \sqrt{m}\, \omega} = x_0^2 \;,
\eea
where we used \eqref{semicldff}.
The width $\Delta x$ of $|\psi_0|^2$ around $x_p\approx x_0$ is determined by demanding
\bea
e^{- \frac{m \omega}{ \hbar} (\Delta x)^2} = e^{-1} \;,
\eea
which yields
\bea
(\Delta x)^2 = \frac{\hbar}{m \omega} = \frac{x_0^2}{2 r_0} \:.
\eea
Thus, in the semi-classical limit, 
\bea
\frac{|\Delta x|}{x_0} = \frac{1}{\sqrt{2 r_0}} \rightarrow 0 \;,
\eea
which shows that the groundstate wave function $\psi_0$ is sharply localized at $x_0$.

While in the semi-classical limit we have $2 r_0 = 1+ a \rightarrow \infty$ (cf. \eqref{rocoup}), in the quantum theory we keep
$2 r_0 = 1 + a$ fixed, with $a$ denoting the index of the modified Bessel function of the first kind $I_a$ which determines the  Euclidean heat kernel $K_E (x,y, \beta)$ of the DFF$_\omega$ model, to which we turn in the next section.

\section{The Segal-Bargmann heat kernel in the DFF$_\omega$ model 
\label{sec:heat}}

The heat kernel $K$ of the DFF$_\omega$ model
was computed exactly in Minkowski time
in \cite{khandelawande} by means of a path integral representation.
When Wick rotated to Euclidean time, it takes the following form 
for paths starting at $x$ at time $t_i$ and ending at $y$ at time $t_f$ \cite{LopesCardoso:2025azr},
\begin{equation}
\label{heatKE}
K_E (x,y, \beta)
=
\frac{m \omega\sqrt{xy}}{\hbar \sinh(\omega\beta)}
\exp\left[
-\frac{m \omega}{2 \hbar }(x^2+y^2)\coth(\omega\beta)
\right]
I_a \left(
\frac{m \omega xy}{\hbar \sinh(\omega\beta)}
\right),
\end{equation}
where $\beta = t_f - t_i$,
and \(I_a\) denotes a modified Bessel function of the first kind,
with index $a$ given by \eqref{rocoup}.

In \cite{LopesCardoso:2025azr} it was shown that when evaluated
on a periodic path in configuration space that starts and ends at $x_0$ (with $x_0$ given by \eqref{x0g}),
the diagonal Euclidean heat kernel $K_E (x_0,x_0, \beta)$
encodes, in the limit of small frequency $\omega$, all the power law suppressed corrections to the BPS black hole entropy that appear in the Rademacher series expansion of the exact BPS black hole degeneracies. The resulting heat kernel was denoted by 
${\tilde K}_E (z,a)$ in \cite{LopesCardoso:2025azr}. The latter
 captures neither the exponential growth of the BPS entropy nor the logarithmic corrections to it.

To remedy this, we will now consider paths between specific complexified endpoints, beginning at $(x_0,0)$ and ending at $(0,x_0) $. Accordingly, in \eqref{heatKE} we set
\begin{equation}
x=x_0 \;\;\;,\;\;\;
y=i x_0.
\end{equation}
Then
\begin{equation}
xy=i x_0^2,
\end{equation}
and
\begin{equation}
x^2+y^2
=
x_0^2+(i x_0)^2
=
x_0^2-x_0^2
=
0.
\end{equation}
Therefore the Gaussian factor in \eqref{heatKE} cancels,
\begin{equation}
\exp\left[
-\frac{\omega}{2}(x^2+y^2)\coth(\omega\beta)
\right]
=
1,
\end{equation}
and we obtain\footnote{The resulting heat kernel is of the Segal-Bargmann type. The Segal-Bargmann heat kernel is obtained from $K_E(x,y,\beta)$ by analytically continuing $x$ or $y$ to complex values.}
\begin{equation}
K_E (x_0, i x_0, \beta)
=
\frac{m \omega\sqrt{i}\,x_0}{\hbar \sinh(\omega\beta)}
I_a\left(
i\frac{m \omega x_0^2}{\hbar \sinh(\omega\beta)}
\right).
\end{equation}
Choosing the principal branch
\begin{equation}
\sqrt{i}=e^{i\pi/4},
\end{equation}
we obtain
\begin{equation}
K_E (x_0, i x_0, \beta)
=
e^{i\pi/4}
\frac{m \omega x_0}{ \hbar \sinh(\omega\beta)}
I_a\left(
i\frac{m \omega x_0^2}{\hbar \sinh(\omega\beta)}
\right) \;.
\end{equation}

In the limit \(\omega\to 0\), this reduces to 
\begin{equation}
K_E (x_0, i x_0, \beta)
=
e^{i\pi/4}
\frac{mx_0}{\hbar \beta}
I_a\left(
i\frac{mx_0^2}{\hbar \beta}
\right) \;.
\end{equation}
Now we observe that if we also
reverse Wick rotate to Minkowski time,  so that $\beta \rightarrow i \beta$, we get
\begin{equation}
\label{heatseg}
K (x_0, i x_0, \beta)
=
e^{i\pi/4} \sqrt{\frac{m}{\hbar \beta}} \, \sqrt{z} \,
I_a \left(z\right) \;\;\;,\;\;\; z = \frac{m x_0^2}{\hbar \beta} = \frac{\sqrt{2mg}}{\hbar \omega \beta}\;.
\end{equation}

Next, consider the unnormalized ground state wave function \eqref{normgroundw}, which we denote by
\bea
\phi_0 (x) = 
x^{2 r_0 - \frac12} \, e^{- \frac{m \omega}{2 \hbar} x^2} \;.
\label{unormgroundw}
\eea
Using this, we compute the unnormalized ground state two-point function
\bea
G(x_0, ix_0) \equiv \phi^*_0 (x_0) \,  \phi_0 (i x_0) = e^{i \frac{\pi}{2} ( a + \frac12) } \, x_0^{2(a + \frac12)} = 
e^{i \frac{\pi}{2} ( a + \frac12) } \; \left( \frac{\hbar \beta}{m} \right)^{a + \frac12} \, z^{a + \frac12}
\;,
\eea
where we used \eqref{rocoup}. We then divide \eqref{heatseg}
by $G(x_0, ix_0) $ to obtain the 
the amputated heat kernel
\begin{equation}
\label{amphk}
\boxed{
K^{\rm amp} (x_0, i x_0, \beta) = 
\frac{K (x_0, i x_0, \beta)}{G(x_0, ix_0) }= 
e^{- i \frac{\pi}{2} a  } \, \left( \frac{m}{\hbar }\right)^{a + 1} \; \beta^{-a-1}
z^{-a} \, 
I_a (z) \;.
}
\end{equation}

As we will explain next, 
the combination 
\bea
{\widehat K}^{\rm amp} (x_0, i x_0, \beta)
\equiv e^{ i \frac{\pi}{2} a  } \, \left( \frac{\hbar}{m} \right)^{a + 1} \; K^{\rm amp} (x_0, i x_0, \beta)
=
\beta^{-a-1}
z^{-a} \, 
I_a (z) 
\label{combI}
\eea
is precisely the combination that appears in the exact expression for BPS black hole microstate degeneracies, given as a Rademacher series expansion.

The microstate degeneracies $d(\textgoth{n})$ of 
 1/2 BPS black holes in  $\mathcal{N}=4$ toroidally compactified heterotic string theory and of 1/8 BPS black holes
 in  $\mathcal{N}=8$ toroidally compactified type IIB superstring theory
possess an exact expression in terms of a Rademacher series expansion \cite{Dabholkar:2005dt,Dabholkar:2014ema,Murthy:2015zzy}, which takes the following form (up to an overall numerical constant),
\bea
d(\textgoth{n}) = 
\sum_{\gamma =1}^{\infty} \gamma^{-a-1} \, {\rm KL} (\textgoth{n}, \gamma) \, 
\hat{I}_a (z) \;,
\label{degdn}
\eea
where
\bea
{\hat I}_a (z) = z^{-a } \, I_a (z) \;.
\eea
For large values of $z$, ${\hat I}_a (z) $ behaves as (up to an overall numerical factor)
\bea
{\hat I}_a (z) = e^{ z - (a + \frac12) \ln z} \left( 1 + {\cal O}\left(\frac{1}{z} \right) \right) \;.
\eea

In the case of 1/2 BPS black holes in  $\mathcal{N}=4$ toroidally compactified heterotic string theory, $\textgoth{n}$ denotes the charge bilinear $n$ carried by these black holes,  and their microstate degeneracies $d(n)$ are encoded in 
a modular form of weight $-12$, namely $\eta^{-24}$ \cite{Dabholkar:2005dt}.
In the case of 1/8 BPS black holes in
 in  $\mathcal{N}=8$ toroidally compactified type IIB superstring theory, $\textgoth{n}$ denotes the quartic U-duality invariant charge combination $\Delta$ carried by these black holes,
 and their microstate degeneracies $d(\Delta)$ 
  are encoded in a weak Jacobi form of weight $-2$ and index $1$, namely $\varphi_{-2,1} (\tau, z) = \vartheta_1^2 (\tau,z)/\eta^6(\tau)$ \cite{Maldacena:1999bp,Shih:2005qf}. The combination $\gamma z$ denotes the leading BPS black hole entropy, which for the 
 1/2 BPS black holes equals $\gamma z = 4 \pi \sqrt{n}$, and for
 the  1/8 BPS black holes equals $\gamma z = \pi \sqrt{\Delta}$.
In \eqref{degdn}, $\rm KL$ denotes a Kloosterman sum, which for  1/2 BPS black holes
is the classical Kloosterman sum given by 
\bea
{\rm KL} (n,-1, \gamma) =  
\sum_{\substack{0\leq-\delta<\gamma\\ \alpha\delta = 1 \text{ mod } \gamma}}
e^{2\pi i\left( n\frac{\delta}{\gamma} - \frac{\alpha}{\gamma}\right)},
\label{classKlooster}
\eea
where 
\beq
\begin{pmatrix} - \mathfrak{b}  & - \delta \\ \alpha  & \gamma
\end{pmatrix} \in SL(2,\mathbb{Z}),
\label{modsl2}
\eeq
while for  1/8 BPS black holes it is given by
\bea
{\rm KL} (\Delta, \nu, \gamma)=  \sum_{\substack{0\leq-\delta<\gamma\\ \alpha\delta = 1 \text{ mod } \gamma}}
e^{2\pi i\left( \frac{\Delta}{4} \frac{\delta}{\gamma} \right) } \,
M_{\nu, 1}^{-1} \, e^{2\pi i\left( -\frac{1}{4} \frac{\alpha}{\gamma} \right) } \qquad {\rm with} \quad \nu = \Delta \mod 2 \;,
\eea
where $M_{\mu,\nu}$ is called the multiplier system.
Comparing \eqref{degdn} with \eqref{combI}, we see that the combination $\gamma^{-a-1}\,  \hat{I}_a (z) $ appearing in each summand of the Rademacher series \eqref{degdn} equals the amputated heat kernel \eqref{combI} 
upon the identification 
$\beta = \gamma$ \cite{LopesCardoso:2025azr}.

Let us summarize the relation between the CQM data $(g, \omega, \beta)$ and BPS black hole data $(a, z, \gamma)$:
\bea
\boxed{
\frac{8 m g}{\hbar^2} = 4 a^2 -1 \;\;\;,\;\;\;
\omega = \frac{\sqrt{a^2 - \frac14}}{\gamma z}\;\;\;,\;\;\;
\beta = \gamma \;.
}
\eea
Note that taking the leading BPS black hole entropy $\gamma z$ to be large implies that $\omega$ is small.

Thus, we have reproduced each term in the Rademacher series expansion - up to the Kloosterman sums - via an amputated heat kernel evaluated along a complexified path!

Next, we turn to the CQM derivation of the Kloosterman sums.
This will be done in the next section. We will thus have arrived
at a holographic CQM derivation of the Rademacher series that yields the exact description of the BPS microstate degeneracies!

Subsequently, we will proceed to establish the emergence of the near-horizon AdS$_2$ bulk dual from this CQM.

\section{Kloosterman sums and conformal quantum mechanics \label{sec:kloo}}
\subsection{Classical Kloosterman sums}
In order to reproduce the classical Kloosterman sum \eqref{classKlooster}, we add to the Euclidean action a Wilson line of the form $- i q \int_0^{\beta} A_{\varphi} d\varphi$, where $q$ is a U(1) charge carried by the DFF$_\omega$ particle w.r.t. the flat connection $A_{\varphi}$. The $\gamma$ sector in the Rademacher sum corresponds to a $\mathbb{Z}_{\gamma}$ orbifold of the Euclidean bulk AdS$_2$, which, in the context of the black hole co-ordinates, corresponds to going from \cite{Banerjee:2008ky}
\beq
ds^2 = (r^2-1) d\varphi^2 + \frac{dr^2}{r^2-1} \;\;\;,\;\;\; (r>1,\,\varphi \rightarrow \varphi+ 2\pi),
\eeq to 
\beq
ds^2 = (r^2-\frac{1}{\gamma^2}) d\varphi^2 + \frac{dr^2}{r^2-\frac{1}{\gamma^2}} \;\;\;,\;\;\; (r>1,\,\varphi \rightarrow \varphi+ 2\pi),
\eeq or equivalently, $r \rightarrow \gamma r$ and $\varphi \rightarrow \frac{\varphi}{\gamma}$. As we will see in the next section, $r$ represents the continuum limit of the discrete energy scale index $n$ in the CQM, while the holographic energy scale transforms as $E \rightarrow \gamma E$, consistent with the boundary time scaling. This, in turn,  implies extending the boundary time integral by a factor of $\gamma$ and hence setting $\beta = \gamma $ as in \cite{LopesCardoso:2025azr} (cf. Section \ref{sec:heat}).
The Wilson line integral evaluated over the unit $\varphi$ circle then contributes a holonomy of the form 
$ 2\pi i q \frac{\delta}{\gamma}$, with $0\leq -\delta < \gamma$. Over the extended range of $\gamma$, its net contribute is $ 2\pi i q \delta$. We now add a 'magnetic' charge $q_m$ to the DFF$_\omega$ particle w.r.t. another flat connection $\tilde{A}_{\varphi}$, i.e. we add
 $+ i q_m \int_0^{\beta} {\tilde A}_{\varphi} d\varphi$.
 This latter connection and charge is magnetic in the sense that we demand that it is symplectically dual to the electric sector. This implies its holonomy contribution will be $ {\tilde A}_{\varphi} = 2 \pi i
 \frac{\alpha}{\gamma} \gamma$, where $\alpha\delta = 1 \text{ mod } \gamma$.
Hence, the net Wilson line contribution to the Euclidean heat kernel will then be a phase factor, 
$e^{2 \pi i (q \delta + q_m \alpha )}$. 
Choosing\footnote{This is motivated by the knowledge that in the full 10D string theory background, the charges associated with the BPS solution represent quantised fluxes through dual cycles in the compactified geometry and the $\mathbb{Z}_{\gamma}$ orbifold is accompained by a $\frac{1}{\gamma}$ twist along some of these cycles, resulting in a corresponding fractional quantisation of the charges as in \cite{Sen:2009gy}.}  $q=n/\gamma $ and $q_m = -1/\gamma$ gives us the required classical Kloosterman sum. 
In sum, the net Euclidean action that generates each Rademacher summand in the $1/\eta^{24}$ case takes the form,
\bea
S_E = \int_0^{\beta} dt \left( \frac12 m \dot{x}^2
+ \frac{g}{x^2}+ \frac{1}{2} m \omega^2 x^2 - i ( q A_t -  q_m \tilde{A}_t) \right).
\eea 
We will refer to the above model from now on as the DFF$_\omega$ model for nomenclatural simplicity.

\subsection{Generalised Kloosterman sums}

In the case of the weak Jacobi form of weight $-2$ and unit index, $\vartheta_1^2 (\tau,z)/\eta^6 (\tau)$, each Rademacher summand corresponding to a specific $\gamma$ additionally contains a multiplier system $M_{\nu,1}$, 
given by \cite{Dabholkar:2014ema}
\bea
\label{multeps}
M^{-1}_{\nu,1}
=
C
\sum_{\epsilon=\pm 1}
\sum_{n=0}^{|\gamma|-1}
\epsilon\,
\exp\left[
\frac{i\pi}{2r \gamma}
\left(
\delta (\nu+1)^2
-2(\nu+1)(2rn+2\epsilon)
+\alpha (2rn+2\epsilon)^2
\right)
\right], 
\eea
where
\bea
\label{normC}
C
=
\frac{i\,\operatorname{sgn}(\gamma)}{\sqrt{2r|\gamma|}}\,
\exp\left[
-\frac{i\pi}{4}
-\frac{i\pi}{6}\Phi(\alpha, \gamma, \delta)
\right],
\eea
$\Phi$ is the Rademacher phi function, and level $r = 3$.

The resulting Kloosterman sum is referred to as the generalised Kloosterman sum. The classical component arises as above from Wilson line holonomies. We therefore have to deduce a Hamiltonian origin for this multiplier system. 
In order to do so, we tensor the DFF$_\omega$ system with an additional Hilbert space arising from the quantisation of a 2D phase space with variables $Q$ and $P$ defining the canonically conjugate variables. Thus, we define 
\bea
Q|x\rangle=x|x\rangle,
\qquad
P=-i\frac{\partial}{\partial x},
\qquad
[Q,P]=i.
\eea

We now define the Hamiltonian 
\bea
H_0
=
\frac{r}{2\pi}P^2.
\eea
Then, the corresponding unitary evolution over time \(\gamma\) is generated  by 
\bea
e^{-i\,\gamma H_0}
=
e^{-i\,\gamma\,\frac{r}{2\pi}P^2}.
\eea
Its free-particle heat kernel, with reverse Wick-rotated time, is 
\bea
\label{fphk}
\langle y|e^{-i\gamma H_0}|x\rangle
=
N_\gamma\,
e^{
\frac{\pi i}{2r\gamma}(y-x)^2
},
\eea
with
\bea
N_\gamma
=
\frac{e^{-i\pi/4}}{\sqrt{2r\gamma}}
\qquad
(\gamma>0).
\eea

Next, we include additional boundary terms in the Hamiltonian at both $t=0$ and $t=\gamma$ in the form of locally supported quadratic potential terms, 
\bea
\frac{\pi }{2r\gamma}(\delta-1)Q^2 \delta(t-\gamma) +
\frac{\pi }{2r\gamma}(\alpha-1)Q^2 \delta(t)
.
\eea
The corresponding time evolution operator expectation value between states $|x\rangle$ and $|y\rangle$ is
\bea
\langle y|U|x\rangle
&=&
N_\gamma
e^{
\frac{\pi i}{2r\gamma}(\delta-1)y^2
}
e^{
\frac{\pi i}{2r\gamma}(y-x)^2
}
e^{
\frac{\pi i}{2r\gamma}(\alpha-1)x^2
}
\nonumber\\
&=&
N_\gamma \, 
e^{
\frac{\pi i}{2r\gamma}
\left(
\alpha x^2-2xy+\delta y^2
\right)
}.
\eea
Thus, the unnormalised heat kernel (with reverse Wick-rotated time) is
\bea
\widetilde K^{(\gamma)}(y,x)
=
e^{
\frac{\pi i}{2r\gamma}
\left(
\alpha x^2-2xy+\delta y^2
\right)
}.
\eea

We now relate this to the multiplier system \eqref{multeps} by setting
\bea
y=\nu+1,
\qquad
x=2rn+\epsilon(\mu+1),
\qquad
n=0,\dots,\gamma-1,
\qquad
\epsilon=\pm1 \:,
\eea
with $\gamma \geq 1$.
Then
\bea
\widetilde K^{(\gamma)}
\bigl(
\nu+1,\,
2rn+\epsilon(\mu+1)
\bigr)
=
e^{
\frac{\pi i}{2r\gamma}
\left(
\delta(\nu+1)^2
-2(\nu+1)\bigl(2rn+\epsilon(\mu+1)\bigr)
+\alpha\bigl(2rn+\epsilon(\mu+1)\bigr)^2
\right)
}.
\eea
Therefore, the multiplier system \eqref{multeps} is the antisymmetrised combination
\bea
M^{-1}_{\nu,\mu}
=
C
\sum_{\epsilon=\pm1}
\sum_{n=0}^{\gamma-1}
\epsilon\,
\widetilde K^{(\gamma)}
\bigl(
\nu+1,\,
2rn+\epsilon(\mu+1)
\bigr)
\eea
with $\mu = 1$.
We see that the coordinate $x$ must be chosen to be a compact coordinate such that 
\bea
x\in \mathbb Z/(2r\gamma\mathbb Z).
\eea
Hence, w.r.t the antisymmetrised 
state
\bea
|\mu\rangle_{\gamma,-}
=
\sum_{n=0}^{\gamma-1}
\left(
|2rn+(\mu+1)\rangle
-
|2rn-(\mu+1)\rangle
\right) \;,
\eea
the multiplier matrix element is 
\bea
M^{-1}_{\nu,\mu}
=
C\,
\langle \nu+1|
U
|\mu\rangle_{\gamma,-}.
\eea
Therefore, the Hamiltonian associated with the additional Hilbert space required to generate the phases in the multiplier system is 
\bea
H(t)
=
\,\frac{r}{2\pi}P^2
-
\frac{ \pi }{2r\gamma}(\alpha-1)Q^2\delta(t)
-
\frac{\pi }{2r\gamma}(\delta-1)Q^2\delta(t-\gamma) \;,
\eea
yielding the corresponding evolution operator being,
\bea
U
=
e^{
-i\int_0^\gamma H(t)\,dt
} \;,
\eea
with the endpoint delta-functions generating quadratic potentials with local support.

In Appendix \ref{sec:AppC} we show that in order to reproduce the normalisation pre-factor, we need to include additional time dependent terms in the Hamiltonian, so that 
the full Hamiltonian in $(Q,P)$ reproducing the multiplier system can be written as 
\bea
H(t)
=
\,\frac{r}{2\pi}P^2
-
\frac{\pi }{2r\gamma}(\alpha-1)Q^2\delta(t)
-
\frac{\pi }{2r\gamma}(\delta-1)Q^2\delta(t-\gamma)
+
\left[
\frac{\pi}{6}\varphi (t)
-
\frac{\pi}{2\gamma}
\right] \;.
\eea
Its propagator over one period \(\gamma\) gives the scalar factor
\bea
i\,e^{-i\pi\Phi(\alpha, \gamma, \delta)/6} \;.
\eea
Therefore 
\[
M^{-1}_{\nu,\mu}
=
\sum_{\epsilon=\pm1}
\sum_{n=0}^{\gamma-1}
\epsilon\,
\langle \nu+1|
e^{
-i\int_0^\gamma H(t)\,dt
} \, 
|
2rn+\epsilon(\mu+1)
\rangle.
\]
This is the desired heat-kernel representation of the multiplier system in the ${\mathcal N}=8$ 1/8 BPS black hole degeneracies, written as the Rademacher expansion of Fourier coefficients of $\vartheta_1^2 (\tau, z) /\eta^6(\tau)$.

\section{Krylov basis, continuum limit and emergent AdS$_2$ space-time
\label{sec:Kryl}}

We will now derive the AdS$_2$ bulk geometry that is holographically dual to the DFF$_\omega$ model.

We consider the time evolution of the groundstate $|0\rangle$
with respect to the linear combination of $SL(2, \mathbb{R})$ generators, 
${\cal G} = \omega (L_+ + L_-) + \kappa R $ \cite{Balasubramanian:2022tpr}.
We denote the time evolved state by $|O (t)\rangle$,
\bea
| O(t) \rangle = e^{i {\cal G} t } \, |0 \rangle \;.
\label{timeevolv}
\eea
We expand $| O(t) \rangle$ in the Krylov basis given by $|n\rangle$ ($n \in \mathbb{N}_0$) \cite{Caputa:2021sib}, 
\bea
| O(t) \rangle = \sum_{n=0}^{\infty}  C_n (t) \, |n \rangle \;.
\eea
Differentiating with respect to time gives
\bea
\frac{d}{d t} |  O(t) \rangle = i {\cal G} | O(t) \rangle = \sum_{n=0}^{\infty}  \frac{d}{d t} C_n (t) \, |n \rangle \;.
\eea
Then, using \cite{Caputa:2021sib,Balasubramanian:2022tpr}
\bea
{\cal G} |n\rangle = b_n |n-1\rangle +a_n |n \rangle +  b_{n+1} |n+1 \rangle \;,
\eea
where the Lanczos coefficients are given by (here we set $\hbar = 1$)
\bea
a_n = \kappa \left( n + r_0 \right) \;\;\;,\;\;\;
b_n = \omega \sqrt{n ( n + 2 r_0 -1)} 
\;\;\;, \;\;\; n \in \mathbb{N}_0 \;,
\eea
one obtains 
\bea
-i  \frac{d}{d t} C_n  (t) = b_n \, C_{n-1} (t) + a_n C_n (t)  + b_{n+1} \, C_{n+1} (t) \;\;\;, \;\;\; n \in \mathbb{N}_0 \;
\label{Cd}
\eea
with $C_{-1}(t) = 0$. This is the evolution equation of a wave function on a semi-infinite
1D chain, where the index $n$ labels a chain site, $C_n(t)$ denotes the amplitude on site $n$ and $b_n$ represents the hopping to nearest neighbours. 

Below we will argue that one must choose $\kappa = - 2 \omega$. With this choice, the generator $\cal G$
in \eqref{timeevolv} equals
\bea
{\cal G} = \omega \left( L_+ + L_- -2  R \right) = - \frac{2}{m } H\;, 
\eea
where $H$ is the generator of time evolution with respect to Poincar\'e patch time \cite{LopesCardoso:2025azr}.
Then, 
the solution 
$C_n (t) $ 
of \eqref{Cd}, subject to the boundary conditions $C_0(0) = 1$ and $C_n(0) = 0$ for $n \neq 0$, is
\bea
C_n(t) = \sqrt{\frac{\Gamma(n + 2 r_0)}{\Gamma(2 r_0) \, n!}} \, \frac{ \left( i \omega t \right)^n}{\left( 1 + i \omega t \right)^{n + 2 r_0}} \;\;\;,\;\;\; n \in \mathbb{N}_0 \;.
\eea
The corresponding probabilities are
\bea
|C_n (t)|^2 = \frac{\Gamma(n + 2 r_0)}{\Gamma(2 r_0) \, n!} \, \frac{ \left( \omega^2 t^2 \right)^n}{\left( 1 + \omega^2 t^2 \right)^{n + 2 r_0}} \;\;\;,\;\;\; n \in \mathbb{N}_0 \;.
\eea
Expressing this as
\bea
|C_n(t)|^2 = \frac{\Gamma(n + 2 r_0)}{\Gamma(2 r_0) \, n!} \, (1 - q)^{2 r_0} \, q^n \;\;\;,\;\;\;
q = \frac{\omega^2 t^2 }{1 + \omega^2 t^2} \;\;\;,\;\;\; n \in \mathbb{N}_0 \;,
\eea
we see that $|C_n(t)|^2 $ describes a negative binomial distribution over $n \in \mathbb{N}_0 $.
$|C_n(t)|^2 $ is the probability of being in level $n$. As $t$ increases, $q$ increases towards $1$, and the average level $\langle n \rangle (t) $, also called Krylov complexity \cite{Caputa:2021sib}, grows quadratically in time,
\bea
\langle n \rangle (t)  = \sum_{n=0}^{\infty} n |C_n(t)|^2 = 2 r_0 \frac{q}{1-q} = 2 r_0 \omega^2 t^2 \;.
\label{avern}
\eea

Now we consider the continuum limit of \eqref{avern}. To this end, we 
send the site spacing $\epsilon$ to zero, while also scaling $\omega t$, as follows,
\bea
n &=& \frac{r}{\epsilon} \;\;\;,\;\;\; (\omega t)^2 = \frac{T}{\epsilon} \;,
\label{lattspac}
\eea
where $r \geq 0$ and $T>0$ are kept fixed.  Note that the ratios $r/\epsilon$, $T/\epsilon$ and $r/T$ are dimensionless.
The above scaling of $\omega t$ ensures that the mean value  $\langle r \rangle$
stays finite in the limit $\epsilon \rightarrow 0$, 
\bea
\langle r \rangle (T)  = 2 r_0 T \;.
\label{mear}
\eea
The continuum probability density is obtained as follows. We take $2 r_0 \in \mathbb{N}$, in which case
$\frac{\Gamma(n + 2 r_0)}{ n!} \approx n^{2 r_0-1}$ in the large $n$-limit. The probability $|C_n(t)|^2 $ becomes
expressed as
\bea
P_{r/\epsilon} (T) = \epsilon \, \frac{r^{2 r_0-1}}{\Gamma (2 r_0)} \, \frac{e^{- r/T}}{T^{2 r_0}} \;,
\eea
where we used
\bea
\left( \frac{T}{T + \epsilon} \right)^{r/\epsilon} = \left( 1 + \frac{\epsilon}{T} \right)^{- r/\epsilon} \longrightarrow e^{-r/T} \;.
\eea
The continuum probability density is
\bea
\rho(r, T) = \lim_{\epsilon \rightarrow 0} \frac{P_{r/\epsilon} (T)}{\epsilon} = \frac{r^{2 r_0-1}}{\Gamma (2 r_0)} \, \frac{e^{- r/T}}{T^{2 r_0}} 
\eea
and describes a Gamma distribution with shape parameter $2 r_0$ and scale parameter $T$.
Note that $\rho (r, T)$ is normalized,
\bea
\int_0^{\infty} \rho(r, T)  dr = 1 \;.
\eea
It follows that
\bea
\langle r \rangle (T) = \int_0^{\infty} r \, \rho(r, T) dr = 2 r_0 T \;, 
\eea
in agreement with \eqref{mear}.

Next, let us discuss the continuum limit of the amplitude $C_n(t)$. We begin by writing  $C_n(t)$ in the form 
\bea
C_n(t) = |C_n(t)|\, e^{i \Phi_n (t)} \;\;\;,\;\;\; \Phi_n (t) = \frac{n \pi}{2} - (n + 2 r_0) \arctan \omega t  \mod 2 \pi \;.
\eea
In the continuum limit, the phase becomes\footnote{The phase $\Phi_{r/\epsilon}$ should not be confused with the function $\Phi$ in \eqref{normC}.}
\bea
\Phi_{r/\epsilon} (t) = \frac{r \pi}{2 \epsilon} - \left(\frac{r}{\epsilon} + 2 r_0 \right) \arctan \sqrt{ \frac{T}{\epsilon}}  \mod 2 \pi \;,
\eea
which behaves as
\bea
\Phi_{r/\epsilon} (t) = \frac{r}{\sqrt{\epsilon T} } - r_0 \pi + {\cal O}(\sqrt{\epsilon})  \mod 2 \pi \;.
\eea
Thus, the continuum amplitude has a smooth Gamma distribution envelope, but its phase oscillates infinitely rapidly as $\epsilon \rightarrow 0$.

Let us discuss the continuum amplitude further. To this end, we introduce
\bea
\psi_{\epsilon}(r, T) =  \frac{1}{\sqrt{\epsilon}} \, C_n(t) \;\;\;,\;\;\; r = n \, \epsilon  \;\;\;,\;\;\; T = (\omega t)^2 \, \epsilon \;.
\eea
Inserting this into \eqref{Cd} gives for its right hand side (with $a = 2  r_0 -1 $)
\bea
\sqrt{\epsilon } \left[ \epsilon \omega
\left( ( r \psi_{\epsilon}' ) ' - \frac{a^2}{4 r} \psi_{\epsilon} \right)   + {\cal O} (\epsilon^2) \right] \;,
\eea
where $'$ denotes a derivative with respect to $r$.
Note that there are no divergent terms of the form $(r/\epsilon ) \psi $ in the big bracket; all such terms cancel out by choosing $\kappa = - 2 \omega$, as we did.

Using $\frac{d}{d t } C_n(t) = 2 \omega  \epsilon \sqrt{T}
\frac{d \psi_{\epsilon}}{d T } $ on the left hand side of \eqref{Cd}, we obtain 
\bea
- 2 i   \sqrt{T} \, 
\frac{\partial \psi_{\epsilon}}{\partial  T} = \sqrt{\epsilon} \left[  ( r \psi_{\epsilon}' ) ' - \frac{a^2}{4 r} \psi_{\epsilon} \right] + {\cal O} (\epsilon^{3/2}) \;.
\label{difps}
\eea
Note that $\psi_{\epsilon}$ has a rapidly oscillating phase, which compensates for the factor $\sqrt{\epsilon}$ on the right hand side. Indeed, \eqref{difps} is solved by
\bea
\psi_{\epsilon} (r, T) = \frac{r^{r_0-1/2}}{\sqrt{\Gamma (2 r_0)}} \, \frac{e^{- r/2T}}{T^{r_0}} \, e^{i \left( \frac{r}{\sqrt{\epsilon T} } - r_0 \pi \right)} \;\;\;,\;\;\; \psi_{\epsilon} (r, T = 0) = 0 \;,
\eea
up to $\sqrt{\epsilon}$-terms that vanish in the limit $\epsilon \rightarrow 0$.

Changing variables from $T>0$ back to $t>0$, \eqref{difps} becomes
\bea
-  i \, 
\frac{\partial \psi_{\epsilon} (t,x) }{\partial  t} =
\epsilon \, \omega \, 
\left[  ( r \psi_{\epsilon}' ) ' - \frac{a^2}{4 r} \psi_{\epsilon} \right] + {\cal O} (\epsilon^{2}) \;.
\label{difps2}
\eea
Note that 
demanding that $T$ is kept fixed as $\epsilon \rightarrow 0$ implies that $t$ scales as $1/\sqrt{\epsilon}$. Therefore the continuum UV cut-off $\epsilon$ is correlated to the large $n$ cutoff, which is simultaneously a late time cut-off.

The Schr\"odinger equation \eqref{difps2} has an $SL(2,\mathbb{R})$ symmetry: 
the equation is invariant under 
\bea
&& t \mapsto t' = \frac{a t + b}{c t + d} \;\;\;,\;\;\;
 r \mapsto r' = \frac{r}{(c t + d )^2 }
\;\;\;,\;\;\; a d - b c = 1 \;, \nonumber\\
&& 
\psi_{\epsilon} (t',r') = (c t + d ) \, e^{- i \frac{c r}{C ( c t + d )} } \, \psi_{\epsilon} (t, r) \;,
\label{trtraf}
\eea
where $C = \epsilon \, \omega$, so that we observe that the wave function satisfies 
the modular (but not the elliptic) transformation properties of a Jacobi form $f_{1,m}(\tau =t, \sqrt{r})$ of weight $1$ and index $m = -\frac{1}{2\pi C}$.
The associated generators are 
\bea
{\cal H} = \partial_t \;\;\;,\;\;\; {\cal D} = t \partial_t + r \partial_r + \frac12\;\;\;,\;\;\; {\cal K} = t^2 \partial_t + 2 t r \partial_r + t -
\frac{i}{C}  r \;,
\label{gentx}
\eea
and they satisfy the ${\rm sl} (2, \mathbb{R})$ Lie algebra commutation relations
\bea
[{\cal H}, {\cal D}] = {\cal H} \;\;\;,\;\;\; [{\cal H}, {\cal K}] = 2 {\cal D} \;\;\;,\;\;\; [{\cal D}, {\cal K}] =  {\cal K} \;.
\eea

We geometrise this symmetry group by demanding it to be the isometry group near the boundary $r=0$ of the AdS$_2$ Poincar\'e patch metric
\bea
ds^2 = \frac{-dt^2 + dr^2}{r^2} \;\;\;,\;\;\; r> 0 \;.
\eea 
This metric possesses the finite isometries
\bea
t \mapsto t' &=& \frac{ (a t + b)(c t + d) - a c r^2}{(c t + d)^2 - c^2 r^2} \;, \nonumber\\
r \mapsto r' &=& \frac{r}{(c t + d)^2 - c^2 r^2} \;,
\eea
which near the boundary $r=0$ reduce to \eqref{trtraf}.

Notice that this metric, apart from inheriting the symmetry group of the holographic DFF$_\omega$ model, has a radial $r$ direction that represents the holographic energy scale
of this model, and hence is a candidate for its bulk holographic dual space-time. We therefore identify it with the AdS$_2$ attractor geometry associated with the BPS black hole, which is a solution to a 2D theory of gravity in the presence of a negative cosmological constant, reduced from a parent 4D gravitational theory. This 2D theory is, by itself, a chiral 2D CFT \cite{Strominger:1998yg}. Being a diffeomorphism invariant theory, its on-shell Hamiltonian expectation value is zero, which implies $\langle 0|L_0|0 \rangle = 0 $, where here $|0\rangle$ represents the supersymmetric AdS$_2$ groundstate of the gravitational theory. The generator $R$ in \eqref{gensl}
is the generator of translations in Euclidean black hole time $\varphi$ \cite{LopesCardoso:2025kry}. In Euclidean AdS$_2$, 
when doing a conformal transformation\footnote{The metrics $ds^2 = (dt_E^2 + dr^2)/r^2 $ with $r >0$ and 
$ds^2 = \sinh^2 \rho \, d \varphi^2 + d \rho^2$ with $\rho>0$ and $- \pi < \varphi <  \pi$ are related by $t_E = - \sinh \rho \sin \varphi/(\cosh \rho - \sinh \rho \cos \varphi)$ and $r= 1/ (\cosh \rho - \sinh \rho \cos \varphi)$. Near the boundary $r=0$, which corresponds to $\rho \rightarrow + \infty$, one obtains 
\eqref{tvar}.
} 
from Euclidean Poincar\'e time $t_E$ to 
black hole
time $\varphi$ near the boundary $r=0$,
\bea
t_E = - \cot \frac{\varphi}{2}
\;\;\;,\;\;\; - \pi < \varphi <  \pi \;,
\label{tvar}
\eea
the stress tensor of the chiral 2D CFT
picks up a Schwarzian anomalous term proportional to the central charge, namely 
\bea
\frac{c}{12} \, \{t_E, \varphi\} =  \frac{c}{24}\;,
\label{ctvar}
\eea
which is the ground state expectation value of the operator $R$ in the DFF$_\omega$ model.
Therefore, 
in order to determine this central charge, we 
look at the expectation value $\langle 0|R|0 \rangle $ 
given in \eqref{semicldff}.
Dividing by a factor of $\hbar \omega $ on dimensional grounds,
we see that its numerical value in AdS$_2$ is $\frac{m x_0^2}{2 \hbar  }$. Using the expression for $z$ given in \eqref{heatseg},  this value can also be written as $\frac{z \beta}{2}$, where $z \beta$  denotes the leading order black hole entropy obtained by setting $\beta = \gamma$  \cite{LopesCardoso:2025azr} (cf. Section \ref{sec:heat}).
 This value is equal to the zero-point energy of the holographically dual  chiral 2D CFT, yielding
\begin{equation}
\label{czrel}
   \frac{c}{12}= 
   z \gamma \;.
\end{equation}. 

For the purpose of computing the vacuum entanglement entropy of a finite segment of length $L$
in the chiral 2D CFT, we change variables from $r>0$ to $- \infty < x < + \infty$ by
\bea
r = \lambda \, e^{x/\ell} \;,
\eea
where $\lambda$ and $\ell$ 
denote length scales to ensure correct length dimensionality. 
As we discuss in 
Appendix \ref{sec:appA},
the vacuum entanglement entropy of a finite segment of length $L$ on an infinite line (the $x$-axis)
is ${\cal S}_{\rm ent}= \frac{c}{6} \ln \frac{L}{\epsilon}$ \cite{Holzhey:1994we}, 
which, on using \eqref{czrel}
and transforming to the thermal cylinder, reproduces the leading black hole entropy as the thermal entropy in each $\gamma$-orbifolded sector of AdS$_2$ \cite{Sen:2009gy}.

\section{Conclusions \label{sec:conc}}
The overarching question in quantum gravity has been to arrive at a quantum statistical description of the fundamental degrees of freedom of gravity. This implies a quantum mechanical formulation with a precisely defined Hilbert space of states, which can in a specific coupling regime, become space-time geometries that satisfy the Einstein equations. The AdS/CFT correspondence postulates that this quantum mechanical formulation is holographic and exists in one dimension lower than the gravitational theory. The simplest systems where both holographic emergence of AdS space-times as well as the underlying quantum mechanical formulation can be studied with mathematical rigour are those classes of BPS black holes for which the exact degeneracy formula has been written down in terms of automorphic forms. Further, the entropy or degeneracy of these black holes is conjectured to arise from the near-horizon geometry which contains an AdS$_2$ factor. In these cases, the BPS degeneracy is written as a Rademacher expansion, which is an infinite convergent sum over Ford circles in the hyperbolic upper half-plane. Each summand encodes black hole entropy contributions from the near-horizon AdS$_2$ space-time or its orbifold. In particular, the AdS/CFT correspondence stipulates that the holographic formulation of the near-horizon gravitational dynamics must be a conformal quantum mechanical (CQM) model. This raises the question of how a quantum mechanical model can possibly encode an exponentially large degeneracy corresponding to the Bekenstein-Hawking entropy. Further, how does the near-horizon geometry arise from quantum mechanics? As these black holes are at zero temperature, the black hole entropy must arise from quantum entanglement, and this must be demonstrable in any viable AdS$_2$/CQM proposal for black hole entropy. 
In \cite{LopesCardoso:2025azr}, the authors took the first step in this direction by demonstrating that in the known BPS counting cases, where the counting formula was expressed as a Rademacher expansion of the Fourier coefficient of a modular or Jacobi form, the terms that captured the power law suppressed
corrections to black hole entropy in each Rademacher summand were precisely encoded in a Euclidean heat kernel calculation in a DFF model coupled to a harmonic oscillator, suggesting its candidature as a model that captured a universal sector of the underlying holographic CQM in these cases. This model was assumed to arise in the large $N$ limit of an $N$-particle Calogero model, which further captured both the leading Bekenstein-Hawking entropy as well as logarithmic corrections to it. However, even though there was now a quantum mechanics proposal for BPS black hole state counting, it was not demonstrated to be the holographic dual to the near-horizon AdS$_2$ attractor geometry, leaving the question of the holographic origins of BPS black hole entropy unanswered.

In this note, we first explicitly demonstrated that the holographic CQM model relevant to black hole counting is, in fact, the large $N$ limit of a Calogero model in a harmonic trap, best formulated in terms of spherical degrees of freedom with the angular sector dynamics frozen (i.e. a fixed angular momentum state) so that the remaining extant mode is a radial mode, whose dynamics is effectively captured by a DFF model in a harmonic trap restricted to zero momentum. 
We then computed the amputated Euclidean heat kernel along complexified paths, and showed that it precisely encodes the Bekenstein-Hawking, the logarithmic corrections as well as all the subleading corrections to the BPS black hole entropy. This is the first known holographic computation of exact BPS black hole entropy in string theory. In order to derive the emergent bulk dual, we turned to the time evolution of the $R$-vacuum in the DFF$_\omega$ model in the discrete Krylov basis, and taking the continuum limit, we obtained a Lanczos-Schr\"odinger equation, which is invariant under a $SL(2,\mathbb{R})$ group of symmetries. Geometrising these symmetries by demanding that this group be the asymptotic isometry group of a 2D space-time, where the sole spatial co-ordinate is the continuum limit of the discrete energy spectrum index, gave rise to a Poincar\'e AdS$_2$ geometry. The radial scale of this geometry is, by construction, the energy scale of the holographic boundary theory, in line with the AdS$_2$/CQM correspondence. This is therefore a candidate bulk dual space-time. Identifying this with the attractor geometry, implies that we embed this in a 2D theory of quantum gravity, which is a chiral 2D CFT \cite{Strominger:1998yg}. Subsequently, we deduced the central charge of this theory, via a conformal transformation at the radial boundary, to black hole co-ordinates
and equating the ground state stress tensor to the ground state energy of the holographic DFF$_\omega$ model. This enabled us to compute the vacuum entanglement entropy in the bulk 2D CFT, from which we extracted the Bekenstein-Hawking entropy by a conformal transformation to the thermal cylinder, demonstrating the encoding of said entropy as an entanglement entropy. We note that this entanglement entropy calculation is performed in the Poincar\'e patch as opposed to previous computations \cite{Azeyanagi:2007bj,LopesCardoso:2025kry} that have been performed in the global AdS$_2$ framework. We also point out that the Krylov basis is the starting point for computing the Krylov complexity in the DFF$_\omega$ model, and the corresponding Fisher information metric over the space of coherent states is precisely the attractor AdS$_2$ background (cf. Appendix \ref{appB}). 
Hence, we have re-packaged exact number-theoretic data in microscopic BPS black hole degeneracy formulae in terms of a conformal quantum mechanics DFF$_\omega$ model, thus reformulating black hole entropy in terms of conformal quantum mechanics and demonstrating the emergence of the near-horizon bulk 
AdS$_2$ space-time. We have thereby inferred both sides of the AdS$_2$/CQM thesaurus from the automorphic structures that govern the organisation of BPS states.

\section*{Acknowledgements}
GLC would like to thank Centro de Ciencias de Benasque Pedro Pascual for hospitality during the completion of this work, and Neil Lambert, James Sparks, Stefan Vandoren and the other participants of the workshop ``Gauge theories, supergravity and superstrings" for discussions.
Work funded by FCT/Portugal and the Recovery and Resilience Plan (PRR) through projects UID/04459/2025 and UID/PRR/04459/2025 (CAMGSD, IST-ID), 
and through project UID/04561/2025 (CEMS.UL).

\appendix

\section{Vacuum entanglement entropy and thermal entropy in a chiral 2D CFT \label{sec:appA}}

In a chiral 2D CFT, the thermal entropy can be extracted from the vacuum entanglement entropy by using a conformal transformation that maps the vacuum on the complex plane to a thermal state on the cylinder, as follows \cite{Caputa:2013lfa}.

Consider a chiral 2D CFT with central charge $c$ on the complex plane $z = x + i y$.  Consider an interval on the $x$-axis of length $L=|z_1-z_2| = |x_1-x_2|$.
The chiral vacuum entanglement entropy of this subsystem on the real line $x$ is, to leading order, 
\bea
{\cal S}_{\rm ent} = \frac{c}{6} \ln \frac{|z_1 - z_2|}{\sqrt{\epsilon_{z_1} \epsilon_{z_2}}}   \;,
\label{entz}
\eea
where $\epsilon_{z_i}$ denote short-distance cutoffs. Now we map the plane to a thermal cylinder with complex coordinate $\sigma$ by
\bea
z = e^{2 \pi \sigma/{\tilde \beta}} \;\;\;,\;\;\; \sigma = u + i \tau \;\;\;,\;\;\; \tau \equiv \tau + {\tilde \beta} \;.
\eea
For convenience, we map the interval on the $x$-axis to an interval at equal Euclidean time $\tau = 0$.
Using
$z_1 = e^{2 \pi u_1/{\tilde \beta}}, z_2 = e^{2 \pi u_2/{\tilde \beta}}$, we obtain (we assume $l \equiv u_2 - u_1 > 0$)
\bea
|z_2 - z_1| = e^{\pi (u_1 + u_2)/{\tilde \beta}} \left( e^{\pi (u_2 - u_1)/{\tilde \beta}} - e^{-\pi (u_2 - u_1)/{\tilde \beta}} \right) = 2  e^{\pi (u_1 + u_2)/{\tilde \beta}} \sinh \left( \frac{\pi l}{\tilde \beta} \right) \;.
\eea
The short-distance cutoffs $\epsilon_{z_i}$ are related to the short-distance cutoff $\varepsilon$ on the cylinder by
\bea
\epsilon_{z_i} = | \frac{dz}{d \sigma} (\sigma_i) | \; \varepsilon = \frac{2 \pi}{{\tilde \beta}} e^{\pi u_i/{\tilde \beta}} \; \varepsilon \;,
\eea
and hence
\bea
\sqrt{\epsilon_{z_1} \epsilon_{z_2}} =  \frac{2 \pi}{\tilde \beta}  e^{\pi (u_1 + u _2)/{\tilde \beta}} \; \varepsilon \;.
\eea
Substituting these expressions into \eqref{entz} yields
\bea
{\cal S}_{\rm ent} = \frac{c}{6} \ln \left( \frac{{\tilde \beta}}{\pi \varepsilon } \sinh \left( \frac{\pi l}{{\tilde \beta}} \right) 
\right)\;.
\label{entsig}
\eea
In the limit $l \gg {\tilde \beta}$,  this becomes
\bea
{\cal S}_{\rm ent} \approx \frac{c}{6} \left( \frac{\pi l}{\tilde \beta} + \ln  \left( \frac{\tilde \beta}{2\pi \varepsilon } 
\right) \right) \;.
\eea
The first term is extensive and yields the thermal entropy as
\bea
{\cal S}_{\rm thermal} = \frac{\partial {\cal S}_{\rm ent}}{\partial l} = \frac{\pi c}{6 {\tilde \beta}}
\;.
\eea
Then, setting 
\bea
{\tilde \beta} = 2 \pi \gamma
\eea
(with $\gamma \in \mathbb{N}$) gives
\bea
{\cal S}_{\rm thermal} = \frac{c}{12 \gamma} = z \;,
\eea
where $z$ denotes the leading entropy in each $\gamma$-orbifolded sector (cf. Section \ref{sec:Kryl}).

\section{Fisher information metric \label{appB}}

We consider the coherent state $|\alpha \rangle $
constructed in \cite{Perelomov:1971bd}, 
\bea
|\alpha \rangle = ( 1 - | \alpha|^2)^{r_0} \,
\sum_{n=0}^{\infty} \alpha^n  \, \sqrt{\frac{ \Gamma (2 r_0 +n) }{n! \, \Gamma( 2 r_0)}} \, | n \rangle \;\:\:,\:\:\: |\alpha| < 1 \:.
\label{cohst}
\eea
Here $\alpha \in \mathbb{C}$ is a complex parameter defined in the unit disc $|\alpha| < 1 $. The state $|\alpha \rangle$ is normalised to one, $\langle \alpha | \alpha \rangle = 1 $.

The expectation value of the generator $R$ with respect to $|\alpha \rangle$ is \cite{Caputa:2021sib}
\bea
\langle \alpha| R | \alpha \rangle
&=& \hbar \, r_0 \left[ 1 + \frac{2|\alpha^2|}{1-|\alpha|^2} \right] = \hbar \, r_0 \, \frac{ 1 + |\alpha^2|}{1-|\alpha|^2} \;.
\eea
The Fubini-Study metric associated with the coherent state $|\alpha \rangle$ is \cite{Caputa:2021sib}
\bea
\langle d \alpha | d\alpha \rangle - \langle d\alpha  | \alpha  \rangle \langle \alpha  | d \alpha  \rangle =  2 r_0 \, \frac{d\alpha d {\bar \alpha} }{(1 - |\alpha|^2)^2} \;\;\;,\;\;\; |\alpha| < 1 \;.
\label{infometric}
\eea
This is an Euclidean AdS$_2$ metric with dimensionless scale $2 r_0$. 
Using 
\bea
2 r_0 =  \omega T \, z \,,
\eea
and dividing \eqref{infometric} by the dimensionless quantity $ \omega T$, we obtain
\bea
ds^2_2 = {\cal S}_{\rm BH} 
\, \frac{d\alpha d {\bar \alpha} }{(1 - |\alpha|^2)^2} \;\;\;,\;\;\; |\alpha| < 1 \;,
\label{ads}
\eea
where $ {\cal S}_{\rm BH} = z $.

\section{Hamiltonian origin of normalisation pre-factor of the multiplier system  \label{sec:AppC}}

We now turn to the normalisation prefactor $C$ in the multiplier system \eqref{multeps},
\bea
C
=
i\,\frac{\operatorname{sgn}(\gamma)}{\sqrt{2r|\gamma|}}\,
e^{-\pi i/4}\,
e^{-\,\frac{i\pi}{6}\Phi(\alpha, \gamma, \delta)} \;.
\eea
For the usual Kloosterman range $\gamma \geq 1$, 
this becomes
\bea
C
=
\frac{e^{i\pi/4}}{\sqrt{2r\gamma}}\,
e^{-\,\frac{i\pi}{6}\Phi(\alpha, \gamma, \delta)}\;.
\eea
Comparing with the heat-kernel normalisation (cf. \eqref{fphk})
\bea
N_\gamma
=
\frac{e^{-i\pi/4}}{\sqrt{2r\gamma}} \;,
\eea
we find
\bea
C
=
i\,e^{-\,\frac{i\pi}{6}\Phi(\alpha, \gamma, \delta)}\,N_\gamma \;.
\eea
Thus, 
since
the heat kernel already \eqref{fphk} includes the normalisation $N_\gamma$, the additional scalar factor required to reproduce the exact multiplier system is
\bea
{\mathcal F} =
i\,e^{-\,\frac{i\pi}{6}\Phi(\alpha, \gamma, \delta)} \;.
\eea

The Rademacher phi function is (cf. \cite{Dabholkar:2014ema})
\bea
\Phi(\alpha, \gamma, \delta)
=
\frac{\alpha+\delta}{\gamma}
-
12\,\operatorname{sgn}(\gamma)\,s(\alpha,|\gamma|) \;.
\eea
For \(\gamma>0\),
\bea
\Phi(\alpha, \gamma, \delta)
=
\frac{\alpha+\delta}{\gamma}
-
12s(\alpha,\gamma) \;,
\eea
where $s(\alpha,\gamma)$ denotes the classical Dedekind sum given by \cite{apostol1990modular}
\bea
s(\alpha,\gamma)
=
\sum_{m=1}^{\gamma-1}
\left(\!\left(\frac{m}{\gamma}\right)\!\right)
\left(\!\left(\frac{\alpha m}{\gamma}\right)\!\right) \;.
\eea
Here, $\left(\!\left( x \right)\!\right)$ denotes the sawtooth function defined by
\bea
\left(\!\left(x\right)\!\right)
=
x-\lfloor x\rfloor-\frac12,
\qquad
x\notin\mathbb Z \;,
\eea
and
\bea
\left(\!\left(x\right)\!\right)=0,
\qquad
x\in\mathbb Z \;.
\eea

In order to reproduce $\Phi (\alpha, \gamma, \delta)$ from a Hamiltonian term, 
we 
define the distribution-valued scalar function 
\bea
\varphi (t)
=
\frac{\alpha+\delta}{\gamma^2}
-
12
\sum_{m=1}^{\gamma-1}
\left(\!\left(\frac{t}{\gamma}\right)\!\right)
\left(\!\left(\frac{\alpha t}{\gamma}\right)\!\right)
\delta(t-m) \;.
\eea
Then
\bea
\begin{aligned}
\int_0^\gamma \varphi (t)\,dt
&=
\frac{\alpha+\delta}{\gamma^2}\,\gamma
-
12
\sum_{m=1}^{\gamma-1}
\left(\!\left(\frac{m}{\gamma}\right)\!\right)
\left(\!\left(\frac{\alpha m}{\gamma}\right)\!\right)
\\
&=
\frac{\alpha+\delta}{\gamma}
-
12s(\alpha,\gamma)
\\
&=
\Phi(\alpha, \gamma, \delta).
\end{aligned}
\eea
Therefore
\bea
\int_0^\gamma \varphi (t)\,dt
=
\Phi(\alpha, \gamma, \delta)
\eea
and
\bea
{\mathcal F} = 
e^{- i \,
\int_0^\gamma \left[
\frac{\pi}{6}\varphi (t)
-
\frac{\pi}{2\gamma}
\right]
\,dt} \;.
\eea

Hence, the full Hamiltonian in $(Q,P)$ reproducing the multiplier system can be written as 
\bea
H(t)
=
\,\frac{r}{2\pi}P^2
-
\frac{\pi }{2r\gamma}(\alpha-1)Q^2\delta(t)
-
\frac{\pi }{2r\gamma}(\delta-1)Q^2\delta(t-\gamma)
+
\left[
\frac{\pi}{6}\varphi (t)
-
\frac{\pi}{2\gamma}
\right] \;.
\eea

\providecommand{\href}[2]{#2}\begingroup\raggedright\endgroup

\end{document}